\documentclass[twocolumn,showpacs,amsmath,amssymb,pra,graphics,graphicx]{revtex4-1}
       \usepackage{graphicx}
        \usepackage{graphics}
\usepackage{bm}
\usepackage{dcolumn}
\begin{document}

\title{Moving Pearl vortices in thin-film superconductors}

\author{V. G. Kogan}
\email{kogan@ameslab.gov}
 \affiliation{Ames Laboratory--DOE, Ames, IA 50011, USA}
 \author{N. Nakagawa}
 \affiliation{Center for Nondestructive Evaluation, Iowa State University, Ames, IA 50011, USA}   

  \date{published in Condens. Matter 2021, 6, 4}
       
\begin{abstract}
The  magnetic field $h_z$ of a moving Pearl vortex in a superconducting thin-film  in $(x,y)$ plane is studied with the help of time-dependent London equation. It is found that for a vortex at the origin moving in $+x$ direction, $h_z(x,y)$ is suppressed in front of the vortex, $x>0$, and  enhanced behind ($x<0$). The distribution asymmetry is proportional to the velocity and to the conductivity of normal quasiparticles. 
  The vortex self-energy and the interaction  of two  moving vortices are evaluated.    
 
 \end{abstract}

\maketitle

\section{Introduction}
 
The time-dependent Ginzburg-Landau equations (GL) are the major tool in modeling  vortex motion. Although this approach, strictly speaking, is applicable only for gapless systems near the critical temperature \cite{Kopnin-Gor'kov}, it  reproduces qualitatively major features of the vortex motion.  

A much simpler  London approach had been successfully employed through the years to describe static or nearly static vortex systems. The London equations express the basic Meissner effect and can be used at any temperature for problems where  vortex cores are irrelevant. The magnetic structure of moving vortices is commonly considered the same as that of a static vortex displaced as a whole.

It has been shown recently, however, that this is not
so for moving vortex-like topological defects in, e.g., neutral superfluids or liquid crystals \cite{leo}.  Also, this is not the  case in superconductors within the Time Dependent London theory (TDL) which takes into account normal currents, a necessary consequence of moving magnetic structure of a vortex \cite{TDL,KP-inter}. In this paper we consider the magnetic field distribution of   moving Pearl vortices in thin films.
We show that the self-energy of a moving vortex decreases with increasing velocity. Moreover, the interaction energy of two vortices moving with the same velocity becomes anisotropic, it is enhanced when the vector $\bm R$ connecting  vortices is parallel to the velocity $\bm v$ and suppressed if  $\bm R \perp \bm v$.
 The magnetic flux carried by moving vortex is equal to flux quantum, but this flux is redistributed so that the part of it in front of the vortex is depleted whereas the part behind it is enhanced.\\


        In   time dependent situations,   the current   consists, in
general, of  normal and superconducting parts:
\begin{equation}
{\bm J}= \sigma {\bm E} -\frac{2e^2 |\Psi|^2}{mc}\, \left( {\bm
A}+\frac{\phi_0}{2\pi}{\bm
\nabla}\chi\right)  \,,\label{current}
\end{equation}
where ${\bm E}$ is the electric field and  $\Psi$ is the order parameter.  

The conductivity  $\sigma$ approaches the normal state value  $\sigma_n$
when the temperature $T$ approaches $T_c$ in fully gapped s-wave
superconductors; it vanishes  fast with decreasing temperature along with the density of normal excitations. This is, however, not the case for strong pair-breaking when   superconductivity becomes gapless while the density of states approaches the normal state value at all temperatures. Alas, there is little experimental information about the $T$ dependence of $\sigma$. Theoretically, this question is still debated, e.g.  Ref.\,\cite{Andreev} discusses possible enhancement of $\sigma$ due to inelastic scattering.

Within the London approach $|\Psi|$ is a constant $ \Psi_0 $ and Eq.\,(\ref{current})
becomes:
\begin{equation}
\frac{4\pi}{c}{\bm J}= \frac{4\pi\sigma}{c} {\bm E} -\frac{1}{\lambda^2}\,
\left( {\bm A}+\frac{\phi_0}{2\pi}{\bm
\nabla}\chi\right)  \,,\label{current1}
\end{equation}
where $\lambda^2=mc^2/8\pi e^2|\Psi_0|^2 $ is the London penetration depth.
Acting on this by curl one obtains:
\begin{equation}
-\nabla^2{\bm h}+\frac{1}{\lambda^2}\,{\bm
h}+\frac{4\pi\sigma}{c^2}\,\frac{\partial {\bm h}}{\partial
t}=\frac{\phi_0}{\lambda^2}{\bm z}\sum_{\nu}\delta({\bm r}-{\bm r_\nu})\,,\label{TDL}
\end{equation}
where ${\bm r_\nu}(t) $ is the position of the $\nu$-th vortex, $\bm z$ is the direction of vortices.
Equation (\ref{TDL}) can be considered as a general form of the time
dependent London equation. 

As with the static London approach, the time dependent version (\ref{TDL}) has the shortcoming of being valid only outside vortex cores.  As such it may produce useful results for materials with large GL parameter $\kappa$ in fields away of the upper critical field $H_{c2}$. On the other hand, Eq.\,(\ref{TDL}) is a useful, albeit approximate, tool for low temperatures where GL theory does not work and the microscopic theory is forbiddingly complex. 

\section{Thin films}

 Let the film of thickness $d$ be in the $xy$ plane. Integration of Eq.\,(\ref{TDL}) over the film  thickness gives for the $z$ component of the field for a Pearl vortex moving with velocity $\bm v$:
\begin{eqnarray}
\frac{2\pi\Lambda}{c}{\rm curl}_z {\bm g} + h_z  +\tau\frac{\partial h_z}{\partial t}=\phi_0 \delta(\bm r -\bm  vt)    .
\label{2D London}
\end{eqnarray}
Here, $\phi_0$ is the flux quantum, $\bm g$ is the sheet current density related to the tangential field components at the upper film face by  $2\pi\bm g/c=\hat{\bm z}\times \bm h$; $\Lambda=2\lambda^2/d$ is the Pearl length, and $\tau=4\pi\sigma\lambda^2/c^2$. With the help of div$\bm h=0$ this equation is transformed to:
\begin{eqnarray}
h_z -\Lambda \frac{\partial h_z}{\partial z}   +\tau\frac{\partial h_z}{\partial t}=\phi_0 \delta(\bm r -\bm  vt) .
\label{hz-eq}   
\end{eqnarray}
 
As was stressed by Pearl \cite{Pearl}, a large contribution to the energy of a vortex in a thin film comes from  stray fields. In fact, the problem of a vortex in a thin film is
reduced to that of the field distribution in free space subject to the boundary condition (\ref{hz-eq}) at the film surface. Since outside the film curl$\bm h=\,\,\,$div$\bm h=0$, one can introduce a scalar potential for the {\it outside} field:
 \begin{eqnarray}
\bm h   =\bm \nabla \varphi,\qquad \nabla^2\varphi=0   \,.
\label{define _phi} 
\end{eqnarray}
The general form of the potential satisfying Laplace equation that vanishes at $z\to\infty$ of the empty upper half-space is 
 \begin{eqnarray}
\varphi (\bm r, z)   =\int \frac{d^2\bm k}{4\pi^2} \varphi(\bm k) e^{i\bm k\cdot\bm r-kz}\,.
\label{gen_sol} 
\end{eqnarray}
Here, $\bm k=(k_x,k_y)$, $\bm r=( x, y)$, and  $ \varphi(\bm k)$  is the two-dimensional Fourier transform of $ \varphi(\bm r, z=0)$. In the lower half-space one has to replace $z\to -z$ in Eq.\,(\ref{gen_sol}). 

As is done in \cite{TDL}, one applies the 2D Fourier transform to Eq.\,(\ref{hz-eq}) to obtain a linear differential equation for $h_{z\bm k}(t)$. Since $h_{z\bm k}=-k\varphi_{\bm k}$, we obtain:
\begin{eqnarray}
\varphi_{\bm k}   =-\frac{\phi_0 e^{-i\bm k\cdot\bm v t}}{k(1+\Lambda k-i \bm k\cdot\bm v \tau) }  \,.
\label{phi(k)} 
\end{eqnarray}
In fact, this gives distributions of all field components outside the film, its surface included. In particular, 
  $h_z$ at $z=+0$ (the upper film face) is given by
\begin{eqnarray}
h_{z\bm k}   =-k\varphi_{\bm k}= \frac{\phi_0 e^{-i\bm k\cdot\bm v t}}{ 1+\Lambda k-i \bm k\cdot\bm v \tau }  \,.
\label{hz(k)} 
\end{eqnarray}
We are interested in the vortex motion  with constant velocity $\bm v=v\hat{\bm x}$, so that we can evaluate this field in real space for the vortex at the origin at $t=0$:
\begin{eqnarray}
h_{z}(\bm r)   = \frac{\phi_0}{4\pi^2} \int \frac{d^2\bm k \,e^{i\bm k\cdot\bm r}}{ 1+\Lambda k-i  k_x v\tau }\,.
\label{hz(r)} 
\end{eqnarray}
 It is convenient in the following to use Pearl $\Lambda$ as the unit length and measure the field in units $\phi_0/4\pi^2\Lambda^2$: 
 \begin{eqnarray}
h_{z}(\bm r)   =  \int \frac{d^2\bm k \,e^{i\bm k\cdot\bm r}}{ 1+ k-i  k_x s }\,,\quad s=\frac{v\tau}{\Lambda}.
\label{hz(r)a} 
\end{eqnarray}
(we left the same  notations for $\bm h_z$ and $\bm k$ in new units; when needed, we   indicate formulas  written in common units). 
 
\subsection{Evaluation of $\bm{h_z(r)}$} 

With the help of identity
\begin{eqnarray}
 ( 1+  k-i  k_x s)^{-1}= \int_0^\infty e^{-u( 1+  k-i  k_x s)} du\,,
\label{ident} 
\end{eqnarray}
one rewrites the field as
\begin{eqnarray}
&& h_{z}(\bm r)   =  \int_0^\infty du\,e^{-u}  \int  d^2\bm k\, e^{i\bm k\cdot\bm \rho-u k  }, \nonumber\\
&&  \bm \rho=(x+us,y) .\qquad
\label{hz(r)1} 
\end{eqnarray}

To evaluate the last integral over $\bm k$, we note that the 3D Coulomb Green's function can be written as 
\begin{eqnarray}
   \frac{1}{4\pi R}=\frac{1}{(2\pi)^3}  \int  \frac{d^3\bm q}{q^2}\, e^{i\bm q\cdot\bm R}  
=  \frac{1}{8\pi^2} \int  \frac{d^2\bm k}{k}\, e^{i\bm k\cdot\bm  r -kz }.\qquad
\label{G(R)} 
\end{eqnarray}
To make here the last step, we used $\bm R=(\bm r,z)$,  $\bm q=(\bm k,q_z)$ and 
\begin{eqnarray}
 \int_{-\infty}^\infty dq_z \frac{ e^{iq_z z}}{k^2+q_z^2}= \frac{\pi\,e^{-k|z|}}{k}\,.
\end{eqnarray} 
 It follows from Eq.\,(\ref{G(R)})
\begin{eqnarray} 
  \int   d^2\bm k \, e^{i\bm k\cdot\bm  r -kz }= -2\pi \frac{\partial}{\partial z} \frac {1}{\sqrt{r^2+z^2}}=\frac{2\pi z}{(r^2+z^2)^{3/2}}.\qquad
\end{eqnarray} 
 Replace now $\bm r\to \bm\rho$, $z\to u$,  $R \to \sqrt{\rho^2+u^2}$ to obtain instead of  Eq.\,(\ref{hz(r)1}):
\begin{eqnarray}
  h_{z}(\bm r)   =2\pi   \int_0^\infty du \frac{u\,e^{-u} }{(\rho^2+u^2)^{3/2}}\,.  
\label{hz(r)d} 
\end{eqnarray}
 After integrating by parts, one obtains:
    \begin{eqnarray}
 h_{z}   =  2\pi  \left[ \frac{1}{r}-  
\int_0^\infty   \frac{du\,e^{-u} }{\sqrt{\rho^2+u^2}} \left(1+\frac{s(x+su)}{\rho^2+u^2} \right)\right] .\qquad
\label{hz(r)2} 
\end{eqnarray}
For the Pearl vortex at rest $s=0$, $ \rho=  r$, and the  known result  follows, see e.g. Ref.\,\cite{jjf}:   
\begin{eqnarray}
 h_{z}(\bm r)   =  2\pi \left\{ \frac{1}{r}+  \frac{\pi}{2} \left [ Y_0(r)-\bm H_0(r)
 \right ]     \right\} \,,
\label{rest} 
\end{eqnarray}
 $Y_0$ and $\bm H_0$ are  second kind Bessel and Struve functions.

Hence, we succeeded in reducing the double integral (\ref{hz(r)a}) to a single integral over $u$. Besides, the singularity at $\bm r=0$ is now explicitly represented by $1/r$, whereas the integral over $u$ is convergent and can be evaluated numerically. 

\begin{figure}[t]
\includegraphics[width=7.5cm] {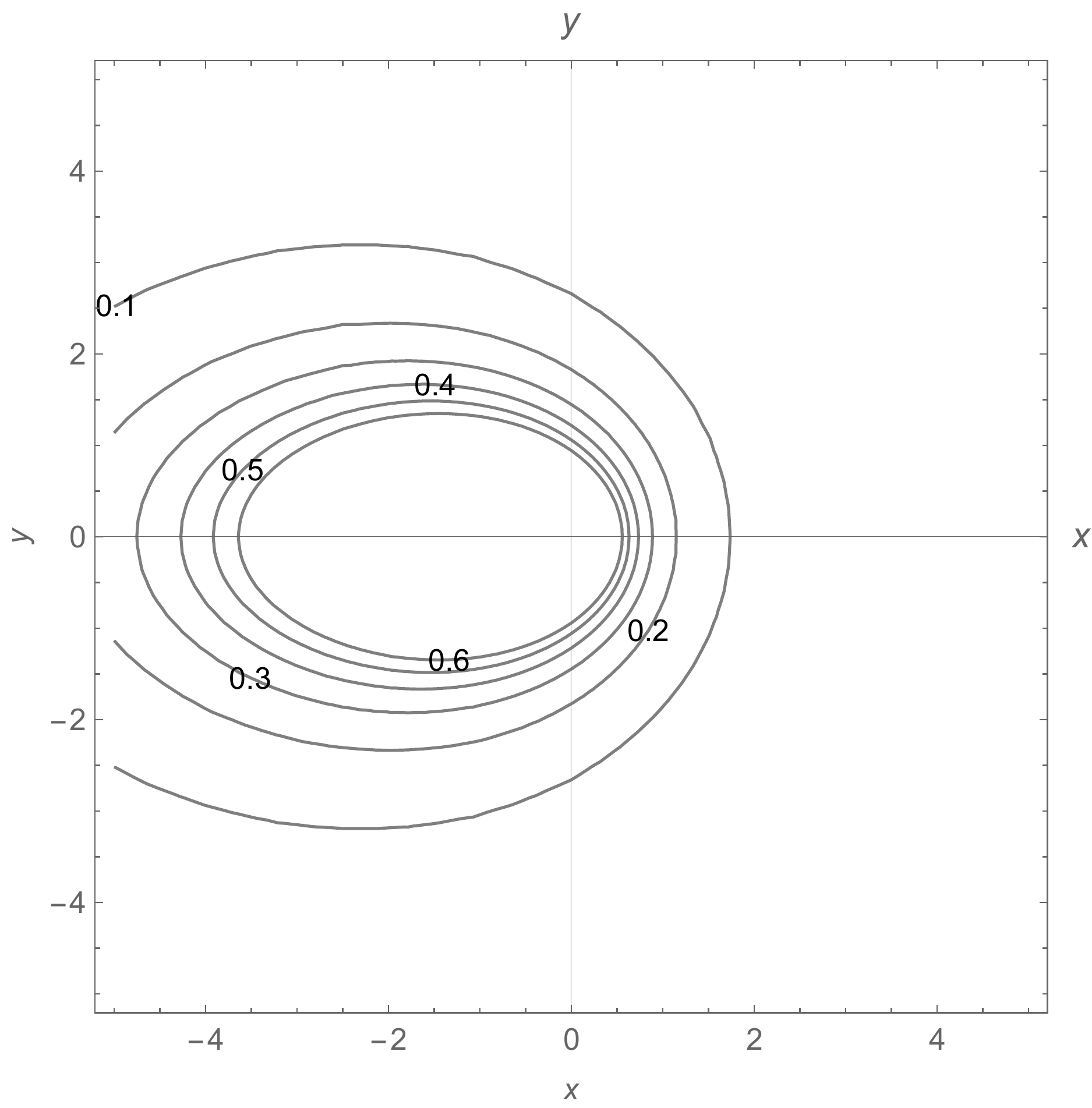}
\includegraphics[width=8cm] {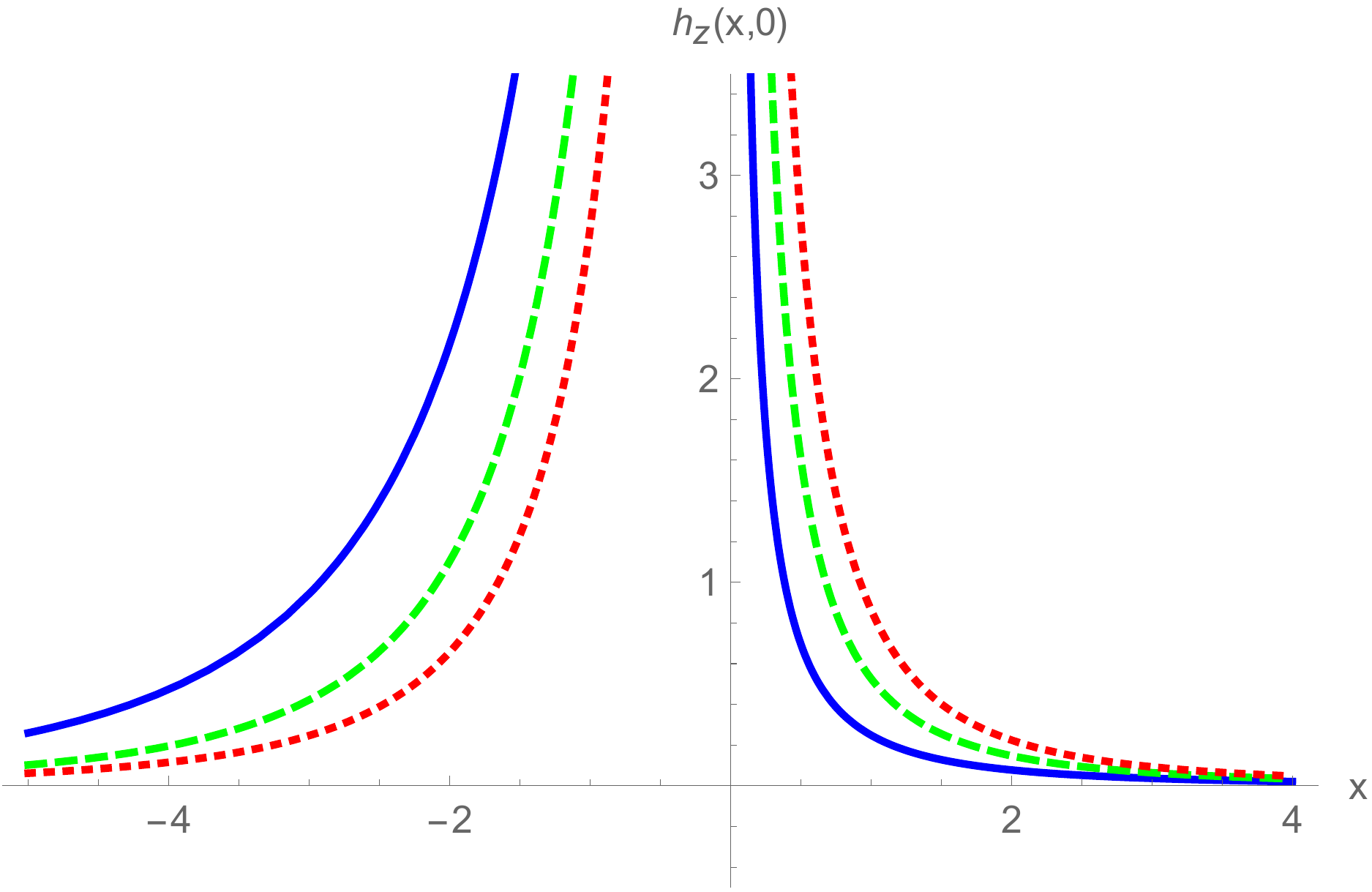}
\caption{  The upper panel: contours of  $h_z(x,y)=\,\,$const ($h_z$ is in units  $\phi_0/4\pi^2\Lambda^2$ and $x,y$ in units of $\Lambda$)  for $s=2$. The lower one: $h_z(x,0)$ for $s=0.5$ (dotted red), $s=1$ (dashed green), and $s=2$ (solid blue).
}
\label{f1}
\end{figure}

The results are shown in Fig.\,\ref{f1}.  
The field distribution is not symmetric relative to the singularity position: the field in front of the moving vortex is suppressed relative to the symmetric distribution of the vortex at rest, whereas behind the vortex it  is enhanced. This is   an interesting consequence of our calculations: the magnetic flux of the moving vortex is redistributed so that it is depleted in front of the vortex and enhanced behind it. 

 We can characterize this redistribution by calculating the magnetic flux $ \Phi_+$ in front of the vortex:
 \begin{eqnarray}  
\frac{ \Phi_+}{\phi_0}&=&    \int_0^\infty dx \int_{-\infty}^\infty dy\, h_z(x,y) \nonumber\\
 &=&    \int_0^\infty dx \int_{-\infty}^\infty dy\int \frac{d^2\bm k}{4\pi^2} \frac{\phi_0 e^{i\bm k \bm r} }{ 1+  k-i   k_xs  }  .
\label{Phi+} 
\end{eqnarray}
The integral over $y$ gives $2\pi\delta(k_y)$, whereas  integrating over $k_x$ we use 
\begin{eqnarray}  
  \int_0^\infty dx\, e^{ik_xx} = i\left(\frac{\cal P}{k_x}-i\pi\delta(k_x)\right) ,
\label{P} 
\end{eqnarray}
where $\cal P$ indicates that the  integral over $k_x$ in Eq.\,(\ref{Phi+}) should be  understood as the principal value. Hence, we have
\begin{eqnarray}  
 \frac{\Phi_+}{\phi_0}=  \frac{i}{2\pi} \left( {\cal P} \int_{-\infty}^\infty  \frac{dk_x  }{ k_x(1+  |k_x|-i   k_xs ) } -i\pi \right).
\label{Phi+a} 
\end{eqnarray}
The integration now is straightforward and we obtain  
\begin{eqnarray}  
 \Phi_+=  \frac{\phi_0}{2} -\frac{\phi_0}{\pi}\arctan s.
\label{Phi+b} 
\end{eqnarray}
 Note that the total flux carried by vortex is given by Fourier component $h_z(\bm k=0)=\phi_0$, see Eq.\,(\ref{hz(k)}), i.e. $\phi_0/2$ is the flux through the   half-plane $x>0$ of the vortex at rest. The flux behind the moving vortex is therefore 
 \begin{eqnarray}  
 \Phi_- =  \frac{\phi_0}{2} +\frac{\phi_0}{\pi}\arctan s.
\label{Phi-} 
\end{eqnarray}
 
\subsection{Potential and London energy of moving vortex}

The potential $\varphi$ introduced above is useful not only as an intermediate step in evaluation of magnetic field, it is directly related to the London energy (the sum of the magnetic energy outside the film and the kinetic energy of the currents inside) \cite{BKT}. 

The potential 
 \begin{eqnarray}  
 \varphi(\bm r )=-\frac{\phi_0 }{4\pi^2\Lambda}   \int \frac{d^2\bm k \,e^{i\bm k\cdot\bm r}}{k( 1+  k-i  k_x s)}.
\label{energy} 
\end{eqnarray}
Employing again the identity (\ref{ident}) we have
 \begin{eqnarray}  
\frac{4\pi^2\Lambda} {\phi_0 } \varphi(\bm r )&=& \int_0^\infty du\,e^{-u}  \int  \frac{d^2\bm k}{k}\, e^{i\bm k\cdot\bm \rho-u k  } \nonumber\\
&=&2\pi \int_0^\infty \frac{du\,e^{-u}}{\sqrt{\rho^2+u^2}}
\label{energy} 
\end{eqnarray}
with $  \rho^2=(x+us)^2+y^2 $.
 
\subsubsection{Self-energy of moving vortex}
 
 This energy    is given by
  \begin{eqnarray}  
\epsilon_0=-\frac{\phi_0}{4\pi}\varphi(\bm r )|_{\bm r\to 0} 
\label{energy1} 
\end{eqnarray}
whereas the integral (\ref{energy}) in this limit is logarithmically divergent. As is commonly done, we can approach the singularity at $\bm r=0$ from any side, e.g. setting $x=0$ and $y=\xi$,   the core size:
  \begin{eqnarray}  
\epsilon_0&=&\frac{\phi_0^2}{8\pi^2\Lambda} \int_0^\infty \frac{du\,e^{-u}}{\sqrt{u^2s^2+\xi_c^2+u^2}}\nonumber\\
&\approx&\frac{\phi_0^2}{8\pi^2\Lambda\sqrt{1+s^2}}  \ln \frac{  \Lambda \sqrt{1+s^2 }}{\xi^2}  
\label{energy2} 
\end{eqnarray}
for the small dimensionless $\xi_c=\xi/\Lambda$. 
 Compare this with the energy of a vortex at rest, see e.g. \cite{BKT}:
 \begin{eqnarray}  
\epsilon_0\approx  \frac{\phi_0^2}{8\pi^2\Lambda} \ln \frac{\Lambda  }{\xi}  \,,
\label{enrest} 
\end{eqnarray}
Hence, the vortex self-energy decreases with increasing velocity, the result qualitatively similar to that of moving vortices in the bulk \cite{KP-inter}.

\subsubsection{Interaction of moving vortices}

\begin{figure}[h ]
\includegraphics[width=7.5cm] {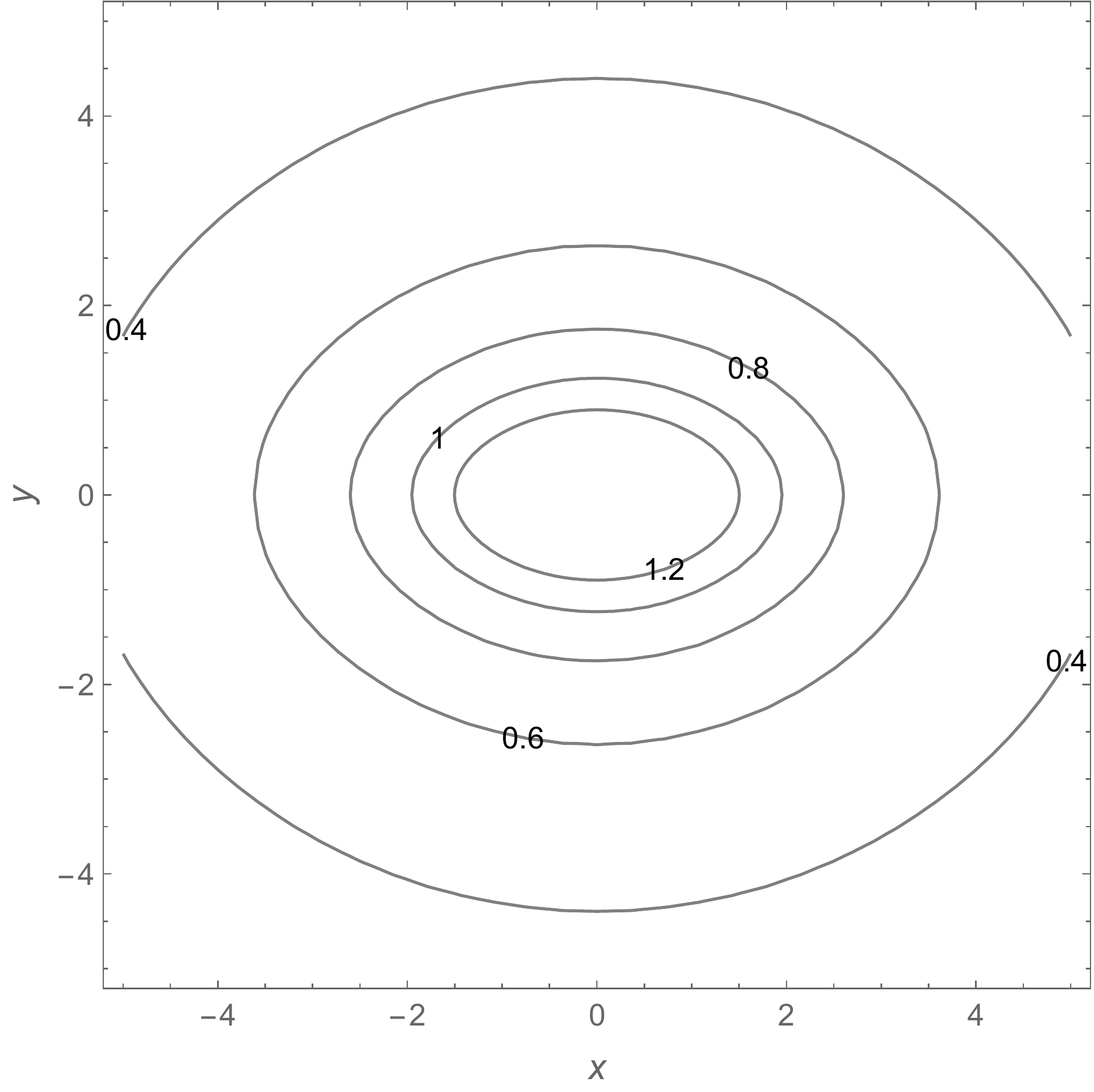}
\caption{(Color online) Contours of constant interaction energy $\epsilon_{int}(x,y) $ for $s=2$.  
}
\label{f2}
\end{figure}

It has been shown in Ref.\,\cite{BKT} that in infinite films the interaction   is given by $\epsilon_{int}=(\phi_0/8\pi)[\varphi_{1}(2) + \varphi_2(1)]$,  $\varphi_1(2)$ is the potential of the vortex at the origin   at the position $\bm r$ of the second. Using Eq.\,(\ref{energy}) we obtain
 \begin{eqnarray}  
\frac{8\pi^2\Lambda}{\phi_0^2 } \epsilon_{int} &=&  \int_0^\infty du\,e^{-u}\Big(\frac{1}{\sqrt{(x+us)^2+y^2+u^2}}\nonumber\\
&+&\frac{1}{\sqrt{(-x+us)^2+y^2+u^2}}\Big)\,.
\label{interact} 
\end{eqnarray}
Clearly, $\epsilon_{int}(x,y)=\epsilon_{int}(-x,y)$.
This energy can be evaluated numerically and the result is shown in 
Fig. \ref{f2} for $s=2$. 
It is worth noting that in thin films the interaction is not proportional to the field of one vortex at the position of the second. In our case the field of one vortex, see Fig.\,\ref{f1}, is not symmetric relative to $x\to -x$, whereas the interaction energy is.

\subsection{Electric field and dissipation}

Having the magnetic field (\ref{hz(k)}) of a moving vortex, one gets for   two vortices, one at the origin and the second at $\bm R$:
\begin{equation}
       h_{z{\bm k}} =  \frac{\phi_0 (1+e^{-i{\bm k} {\bm R} })e^{-ik_xvt}}{1+k\Lambda-i k_x\Lambda s} 
\label{Hk3}
\end{equation}
(in common units). 
The moving nonuniform distribution of the vortex
magnetic field causes an electric field $\bm E$  out of the vortex core, which in turn causes the normal currents  $\sigma\bm E$ and the dissipation $\sigma{\bm E}^2$. Usually this dissipation is small relative to Bardeen-Stephen core dissipation \cite{Bardeen-Stephen}, but for fast vortex motion and high conductivity of normal excitations \cite{Andreev} it can become substantial \cite{TDL}. 
 
\begin{figure}[h ]
\includegraphics[width=7.5cm] {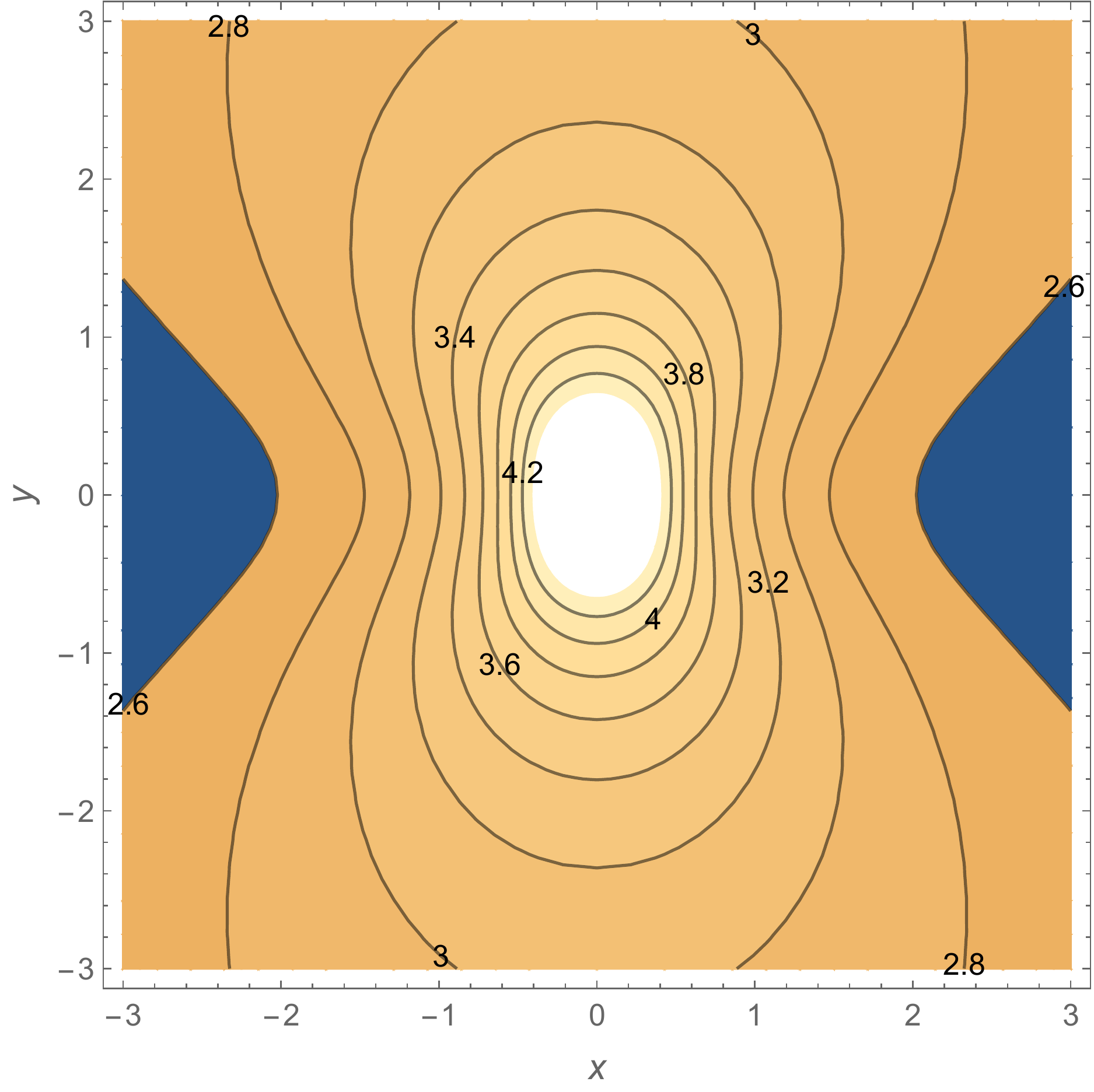}
\caption{  Contours of constant dissipation $w(x,y)$ in the system of two moving vortices for $s=2$. At $t=0$ one is situated    at the origin and the second at $(x,y)$.  }
\label{f3}
\end{figure}

The field $\bm E$ is expressed in terms of known $\bm h$ with the help of the
Maxwell equations $i({\bm k}\times {\bm E}_{\bm k})_z=- \partial_t h_{z{\bm k}}/c$
and ${\bm k}\cdot{\bm E}_{\bm k}=0$:
\begin{eqnarray}
 E_{x{\bm k}} &=&  -\frac{\phi_0v}{c} \frac{k_xk_y(1+e^{-i\bm k \bm R  })}{k^2(
1+k\Lambda-ik_x\Lambda  s )}  \,,\\
E_{y{\bm k}}&=& \frac{\phi_0v}{c} \frac{k_x^2(1+e^{-i\bm k \bm R  })}{k^2(
1+k\Lambda-ik_x\Lambda  s)}  \,.
\end{eqnarray}
For the stationary motion, one can consider the dissipation at $t=0$. 
The dissipation power  is:
\begin{eqnarray}
W=\sigma d\int   d{\bm r}E^2 =\sigma d\int \frac{d^2{\bm
k}}{4\pi^2}\left(|E_{x{\bm k}}|^2+|E_{y{\bm k}}|^2\right)\nonumber\\
=\frac{\phi_0^2\sigma d v^2}{\pi^2c^2} \int d^2{\bm k}\frac{ k_x^2 \cos^2(\bm k \bm R/2)}{k^2[(1+k\Lambda)^2+k_x^2 \Lambda^2s^2]} \,.
\end{eqnarray}
The integral here is divergent at large $k$, but the London theory anyway breaks down in the vortex core of a size $\xi$, so one can introduce a factor $e^{-k^2\xi^2}$ to truncate this divergence. 
We then  calculate the reduced quantity $w(x,y)=W (\pi c^2\Lambda^2/\phi_0^2\sigma dv^2)$ shown in Fig.\,\ref{f3}.
 
An interesting feature of this result is that the dissipation $w(x,y)$  develops a shallow ditch along the $x$ axis. Hence, for a fixed separation of vortices in the pair, the dissipation is minimal if they are aligned along the velocity.

\section{Discussion}

We have shown that in thin films the magnetic structure of the moving Pearl vortex   is distorted relative to the vortex at rest.   The flux quantum of a moving  vortex is redistributed, the back side part of the flux is enhanced, whereas the in-front part is depleted. Physically, the distortion is caused by normal currents arising due to  
changing in time magnetic field at each  point of space, the electric field is induced and causes normal currents. Naturally, it leads to suppression of the flux where it is increasing (in front of the moving vortex) and to enhancement where it is decreasing (behind the vortex).   We characterize this asymmetry  by the difference of fluxes behind ($x<0$) and in front ($x>0$) the moving vortex   $\Delta\Phi=\Phi_--\Phi_+=(2\phi_0/\pi)\arctan s$. For a realistic situation $s=v\tau/\Lambda\ll 1$, although the relaxation time $\tau\propto\sigma\lambda^2$ where $\sigma$ is the poorly-known conductivity of above-the-gap normal excitations.    Measuring $\Delta\Phi$ one can extract $\sigma$, an important physical characteristics of superconductors.
There is an experimental technique which, in principle, could probe the field distribution in moving vortices \cite{Eli}. This is highly sensitive SQUID-on-tip  with the  loop small on the scale of possible Pearl lengths.  

Recent experiments have traced  vortices moving in thin superconducting films with  extremely high velocities well exceeding the speed of sound \cite{Eli,Denis}. Vortices crossing thin-film bridges being pushed by transport currents   have a tendency to form chains directed along the velocity. The spacing of vortices in a chain is usually exceeds by much the core size, so that commonly accepted reason for the chain formation, namely, the  depletion of the order parameter behind moving vortices    is questionable. But at distances $r\gg \xi$ the time dependent London theory is applicable. 

In this paper, we consider only properties of a single vortex and of interaction between two vortices moving with the same velocity,
The problem of interaction in systems of many vortices is still to be considered. 

 \section{Acknowledgements}
The work of V.K. was supported by the U.S. Department of Energy (DOE), Office of Science, Basic Energy Sciences, Materials Science and Engineering Division.  Ames Laboratory  is operated for the U.S. DOE by Iowa State University under contract \# DE-AC02-07CH11358. 

\appendix

\section{ Abrikosov vortex moving in the bulk}

The field distribution of this case has been evaluated numerically in Ref.\,\cite{TDL}.  Here, we provide this distribution in closed analytic form. 

 The magnetic field $\bm h$ has one component $h_z$, so we can omit the subscript $z$. Choosing $\lambda$ as a unit length and measuring the field in units of $\phi_0/4\pi\lambda^2$, we have:
 \begin{equation}
       h({\bm r}) =    \int  \frac{d^2{\bm k}\,e^{ i{\bm k} {\bm r} } }{1+ k^2-i
k_x s}\,,\qquad s=\frac{v\tau}{\lambda}\,.
\label{h}
\end{equation}
First, we use the identity
\begin{eqnarray}
 ( 1+  k^2-i  k_x s)^{-1}= \int_0^\infty e^{-u( 1+  k^2-i  k_x s)} du\,,
\label{identa} 
\end{eqnarray}
so that
\begin{eqnarray}
&& h (\bm r)   = \int_0^\infty du\,e^{-u}  \int  d^2\bm k\, e^{i\bm k\cdot\bm r-u (k^2-ik_xs) } \nonumber\\
&&=  \int_0^\infty du\,e^{-u}  \int  d^2\bm k\, e^{i\bm k\cdot\bm \rho-u k^2  },\,\,\, \bm \rho=(x+us,y) .\qquad
\label{hz(r)1a} 
\end{eqnarray}
Now,  integrals over $k_x,k_y $ are doable: 
\begin{eqnarray}
  \int_{-\infty}^\infty   dk_x\, e^{i k_x \rho_x-u k_x^2  } \int_{-\infty}^\infty   dk_y\, e^{i k_y y-u k_y^2  }=\frac{\pi}{u}e^{-\rho^2/4u}\qquad
\label{int} 
\end{eqnarray}
where $\rho^2=(x+us)^2+y^2$. Hence, we have
 \begin{eqnarray}
  h (\bm r)   &=& \pi\int_0^\infty \frac{du}{u}\,e^{-u-\rho^2/4u}\nonumber\\
  &=&\pi\int_0^\infty \frac{du\,e^{-u}}{u}\,\exp\left[-\frac{(x+us)^2+y^2}{4u}\right] \nonumber\\
  &=& \frac{\phi_0}{2\pi\lambda^2} e^{-sx/2\lambda} K_0\left(\frac{r}{2\lambda}\sqrt{4+s^2}\right) .\qquad
\label{hz(r)1b} 
\end{eqnarray}
the last line is written in common units.
Note that for the vortex at rest $s=0$ and we get the standard result $h=(\phi_0/2\pi\lambda^2) K_0(r/\lambda)$ \cite{deGennes}.

\end{document}